\documentclass[preprint,prd,tightenlines,superscriptaddress,showkeys]{revtex4}

\usepackage{graphicx} 
\usepackage{dcolumn}  
\usepackage{colordvi}
\usepackage{color}
\usepackage{epstopdf}
\usepackage{subfig}
\usepackage{amssymb}
\usepackage{url}
\graphicspath{{ps}}
\usepackage{hyperref}
\usepackage{tabularx}

\input belle2-sym.tex

\def \Y       {\ensuremath{\Upsilon(4S)}\xspace}

\begin{document}


\def\belletwo {{Belle II}\xspace}
\def\itbelletwo {{\it {Belle II}}\xspace}
\def\phaseiii {{Phase III}\xspace}
\def\itphaseiii {{\it {Phase III}}\xspace}

\newcommand\logten{\ensuremath{\log_{10}\;}}

\vspace*{-3\baselineskip}
\resizebox{!}{3cm}{\includegraphics{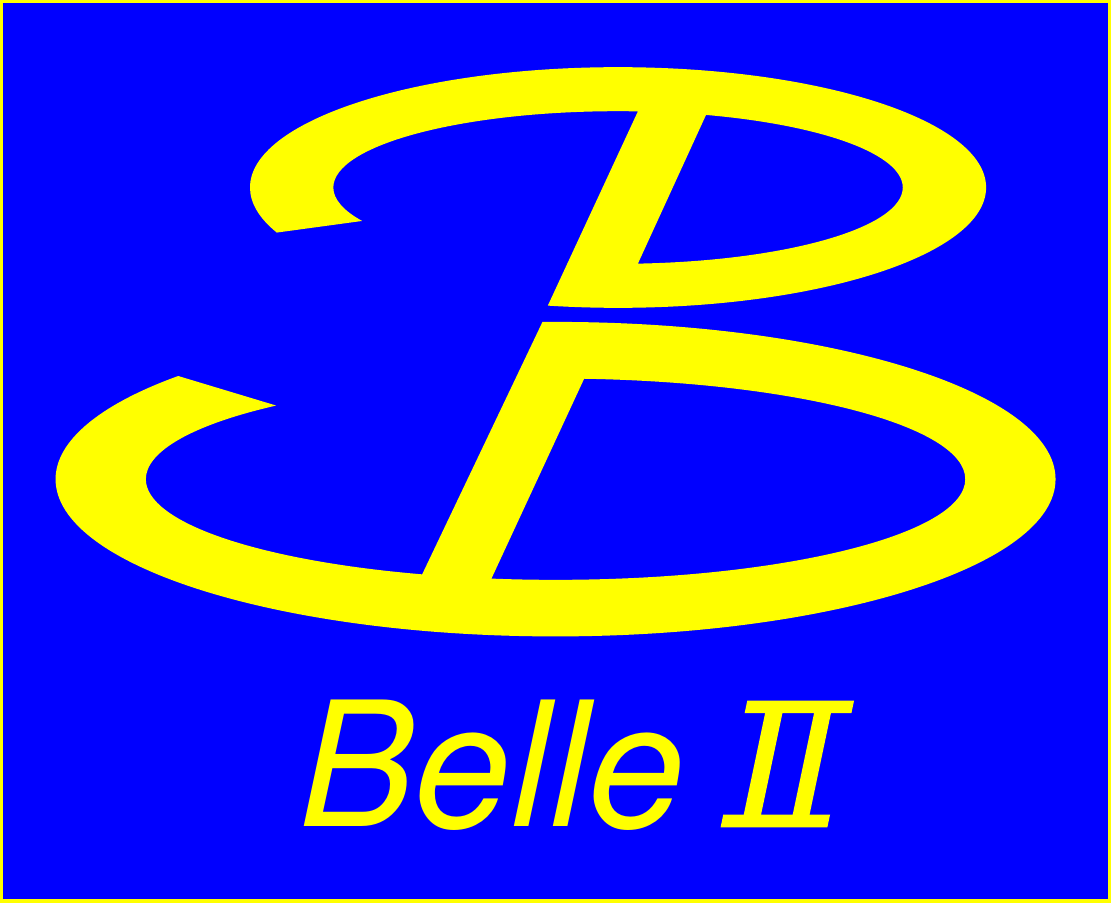}}

\vspace*{-5\baselineskip}
\begin{flushright}
BELLE2-CONF-PROC-2020-014 \\ \today
\end{flushright}

\title { \quad\\[0.5cm] Diversity and inclusion activities in Belle II}
\altaffiliation{\textit{Presented at 40th International Conference on High Energy physics - ICHEP2020; July 28- August 6, 2020; Prague, Czech Republic (virtual meeting)}}

\author{H.M.~Wakeling}
\thanks{Speaker}
\email{hannah.wakeling@physics.mcgill.ca}
\affiliation{Department of Physics, McGill University,\\
3600 rue University, Montr\'eal, Canada}

\author{S.A.~De La Motte}
\email{shanette.delamotte@adelaide.edu.au}
\affiliation{Department of Physics, The University of Adelaide,\\ Adelaide South Australia 5005 Australia}

\author{M.~Barrett}
\email{mattb@post.kek.jp}
\affiliation{Institute of Particle and Nuclear Studies\\
  High Energy Accelerator Research Organization (KEK),\\
  1--1 Oho, Tsukuba, Ibaraki, 305--0801, Japan}

\author{K.~Kinoshita}
\email{kay.kinoshita@desy.de}
\affiliation{University of Cincinnati\\  Cincinnati, Ohio, 45221, U.S.A}

\collaboration{The Belle II Collaboration}
\noaffiliation

\begin{abstract}

These proceedings accompany the Belle II talk in the Diversity and Inclusion parallel session delivered during ICHEP 2020. This marks the first external presentation by the Belle II Collaboration, in which we present some of our data and self-reported statistics regarding diversity and inclusion. We also present Belle II's current and planned activities to aid and improve diversity and inclusion. We find that there is still a lot to be done to improve the social working environment and population representation within our collaboration and within high energy physics.
\end{abstract}

\maketitle

\section*{Introduction}

Belle II is a high energy physics collaboration based in Tsukuba, Japan. The Belle II experiment itself is a particle detector at an electron-positron collider, SuperKEKB. The institutions that contribute to the collaboration span 4 continents. At the end of the 2019 Japanese fiscal year (JFYear) Belle II had 1041 active members registered. This paper will introduce the Belle II collaboration and our efforts to research and improve on diversity and inclusion within the collaboration and outside of it. Section \ref{sec:CurrentClimate} will present the statistics taken from the collaboration's membership registrations and will discuss results from a membership survey taken in 2018. Section \ref{sec:Actions} will exhibit some of the actions Belle II has been taking, and will take, to promote diversity and inclusion.

\section{The current climate at Belle II \label{sec:CurrentClimate}}
\subsection{Belle II collaboration demographics}

This section analyses the demographics of Belle II registration data from the Belle II member registration platforms from 2011 - 2019. In 2017, Belle II migrated from a previous system, maintained by the
Belle II secretariat, to the Belle II Membership Management System (B2MMS), maintained by Belle II Collaborative Services; due to this
migration, results for 2017 may appear anomalous and should not be taken
as indicative of real trends. 
Within the B2MMS, it is compulsory for a member to submit their name, title, gender, email, institution and membership category (i.e. Ph.D. student, faculty, etc.). The possible selections available for gender are: 'Male', 'Female, 'Other', and Unspecified. Here we acknowledge the issues often perpetuated in this form of data taking with respect to a gender binary. 
Additionally, in the following figures it is necessary to combine ‘other’ and ‘unspecified’ into one category as underrepresented persons are easily identifiable due to low statistics. We as a collaboration wish to maintain and pursue true anonymity of our membership data. 
\begin{figure}[h]
    \centering
    \begin{minipage}{.45\textwidth}
        \captionsetup{width=.95\linewidth}
        \includegraphics[width=\linewidth]{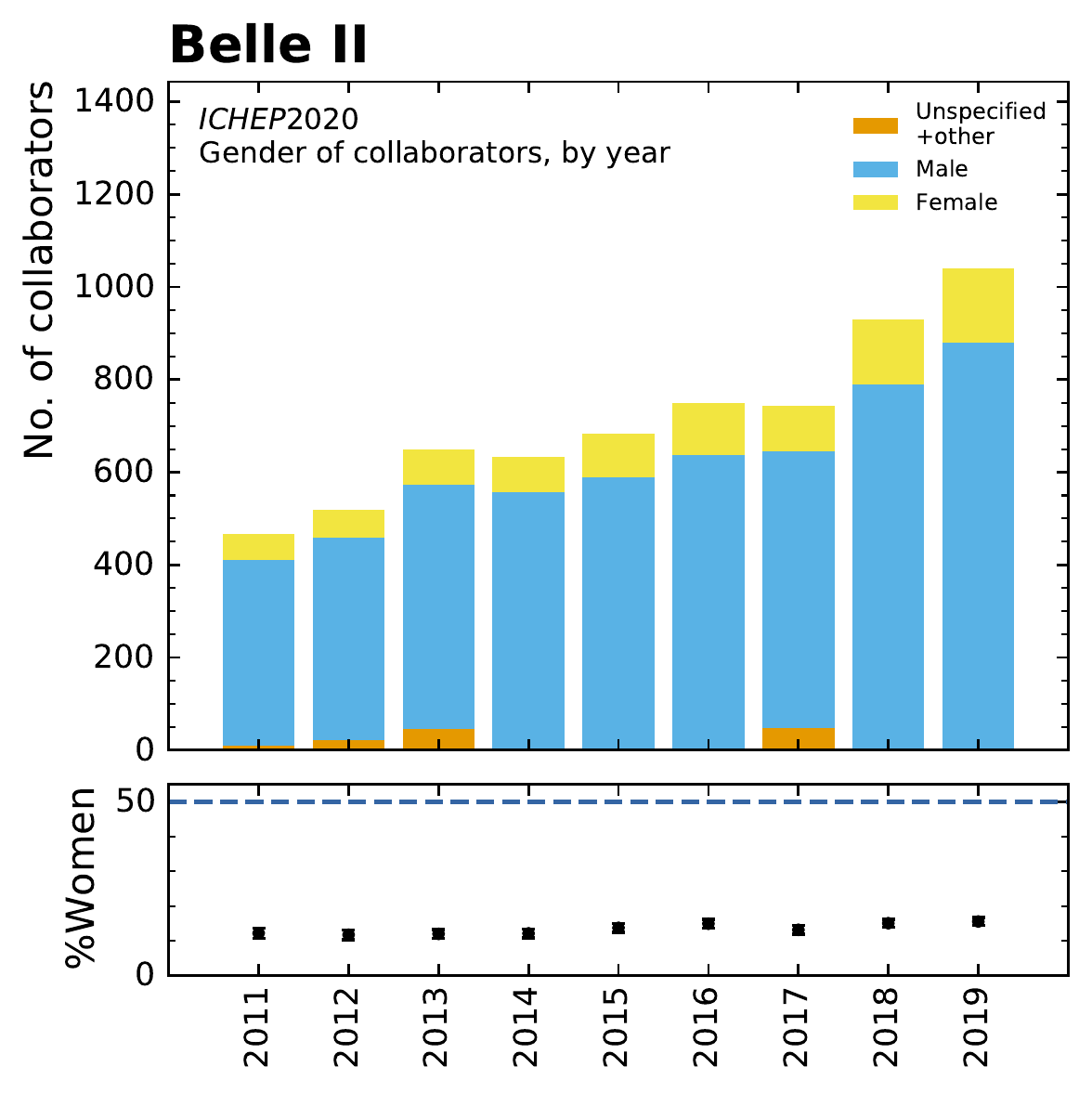}
        \vspace*{1em}
        \caption{The number and gender of Belle II Collaborators, from years 2011 to 2019. Data taken from Belle II membership declaration, where `other' refers to non-binary gender identity and `unspecified' refers to undeclared gender information. \newline \newline}
        \label{fig:by_year}
    \end{minipage}%
    \begin{minipage}{.45\textwidth}
        \captionsetup{width=.95\linewidth}
        \includegraphics[width=\linewidth]{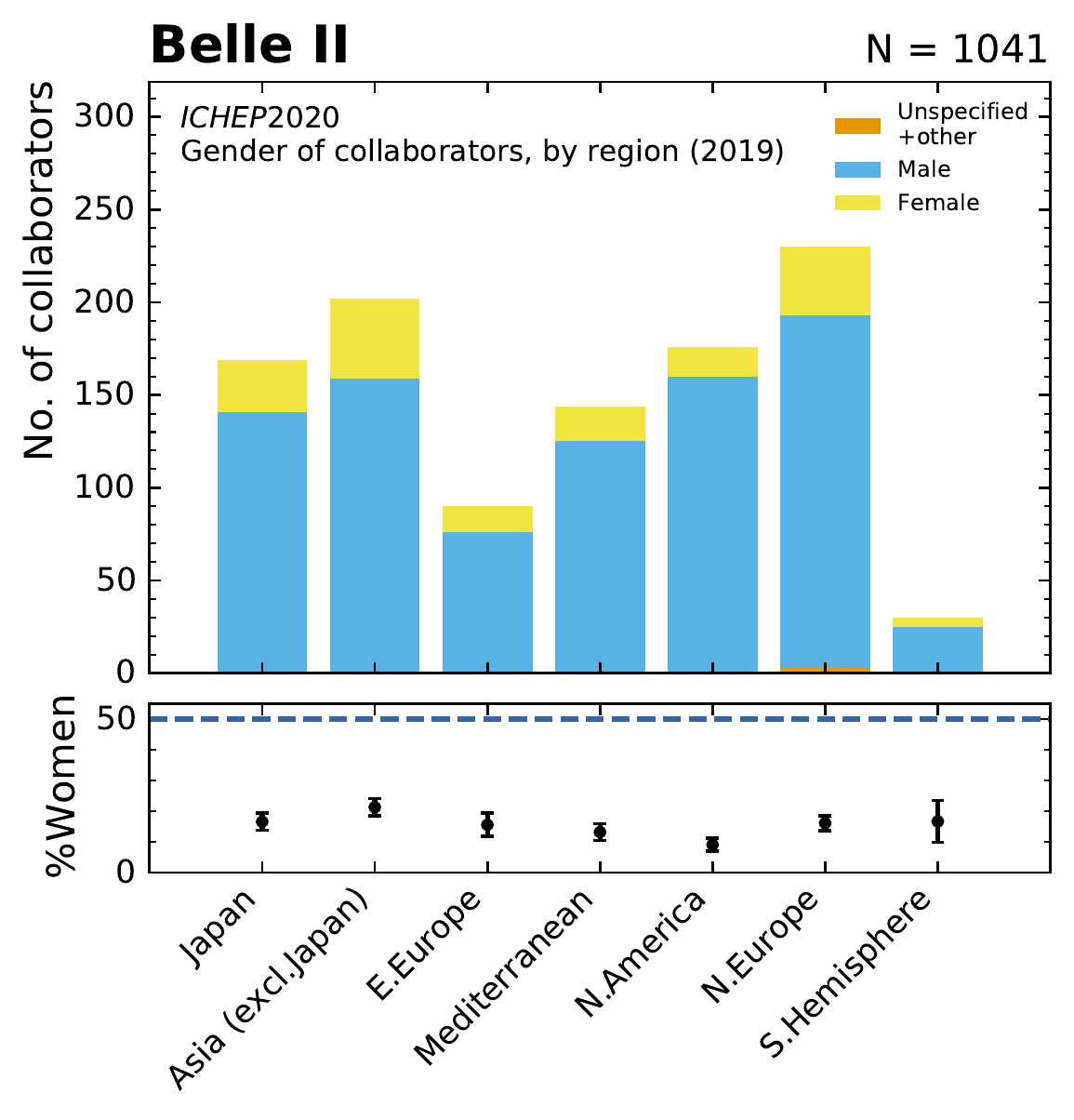}
        \caption{The number of collaborators of the Belle II Collaboration by region and gender. When compared to similar graphs produced by the ATLAS Collaboration \cite{ATL-GEN-PUB-2016-001} we see that we have a much larger proportion of collaborators from Asian institutions than ATLAS.
        Japan alone contributes 16\% of the collaboration, so is separated from the ‘Asia’ region for clearer presentation.}
        \label{fig:by_region}
    \end{minipage}
\end{figure}

In recent years Belle II has seen a distinct membership increase, approximately doubling members between the years 2011 to 2019. Within this, the percentage of female identifying members increased from 12.2\% to 15.6\%, shown in Figure \ref{fig:by_year}. Though statistically significant, the difference of these two numbers shows the increase in the percentage of women members to be approximately 0.4\% every year. Women are significantly underrepresented (equal population representation at 50\% is displayed as a dashed line in the percentage plot of Figure \ref{fig:by_year}). The current rate of increase is too modest to reach gender parity by 2030, when Belle II expects to have reached its target dataset of 50$\,\rm{ab}^{-1}$.

We next compare the gender of collaborators by region they work in, based on their registered institution. One particularly interesting takeaway from this Figure \ref{fig:by_region} is the comparison between two regions or countries with similar representation within the collaboration. Comparing Japan, which represents 16\% of the collaboration, to Northern America which represents 17\% of the collaboration, we can see that Japan has a higher representation of women than Northern America, 17\% compared to 9\% respectively.
\begin{figure}[h]
    \centering
    \begin{minipage}{.5\textwidth}
        \captionsetup{width=.9\linewidth}
        \includegraphics[width=\linewidth]{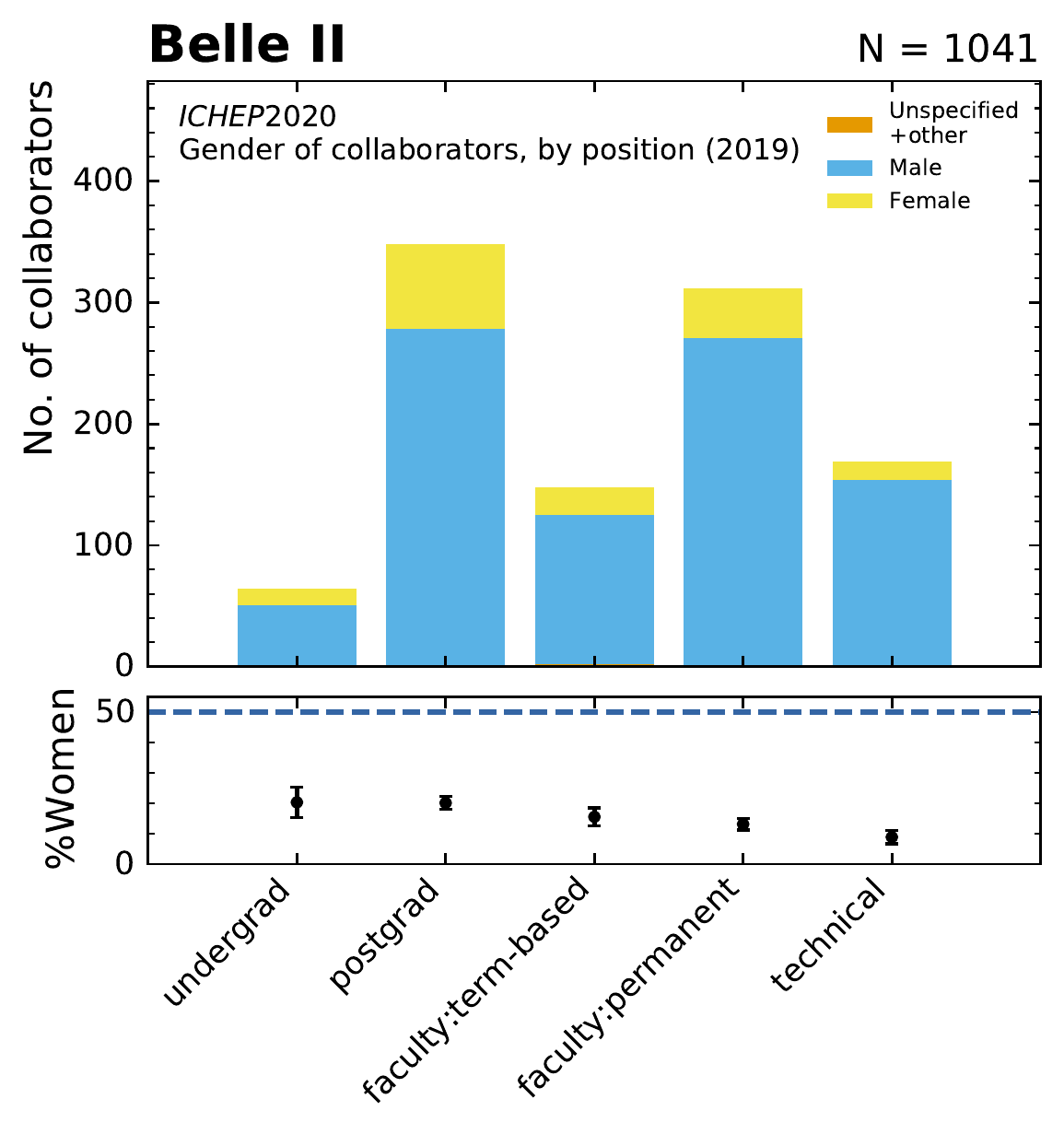}
        \caption{The number of collaborators of the Belle II Collaboration by position and gender.}
        \label{fig:by_pos}
    \end{minipage}%
    \begin{minipage}{.5\textwidth}
        \captionsetup{width=.9\linewidth}
        \includegraphics[width=\linewidth]{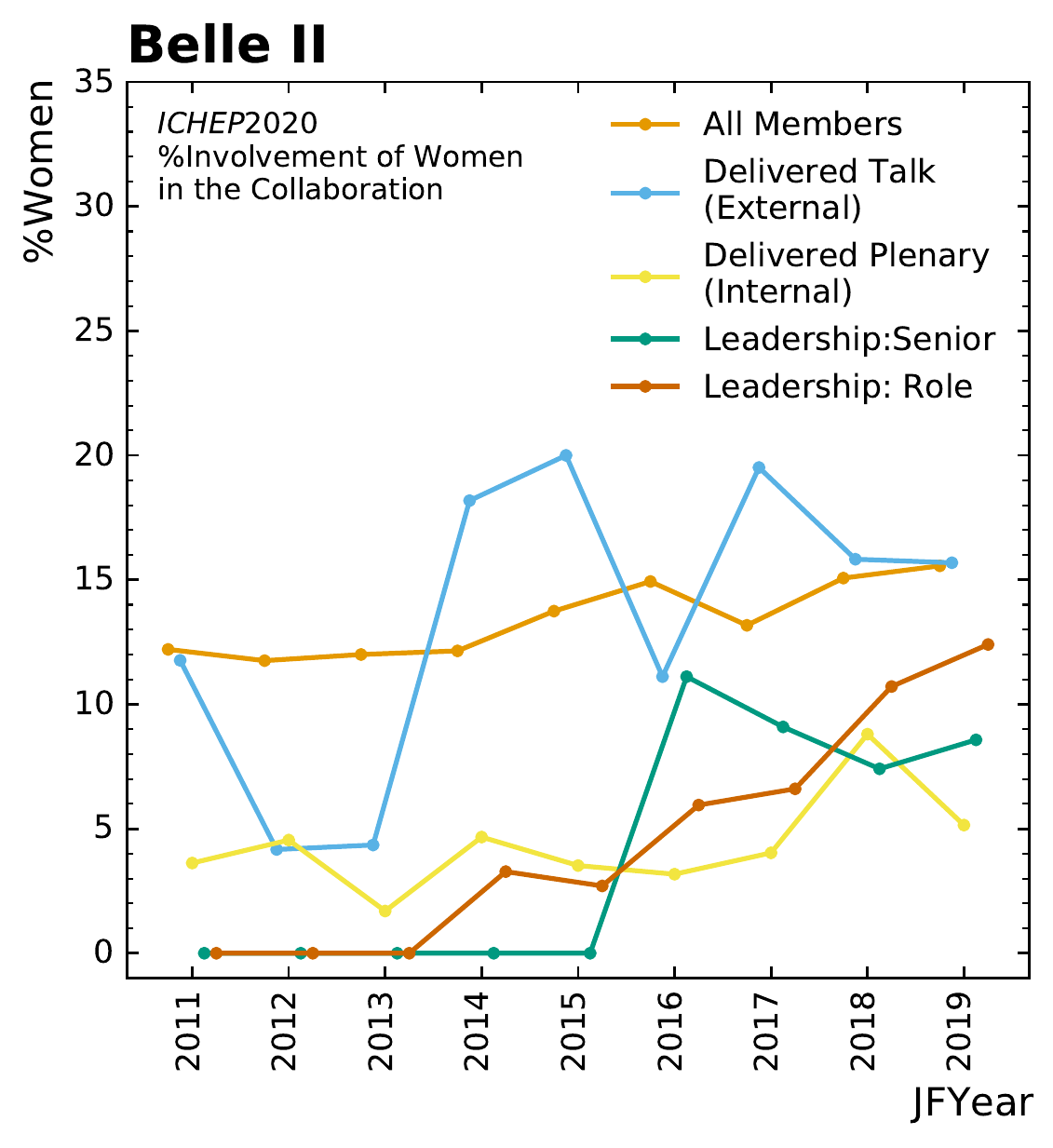}
        \caption{The percentage of people in the specified roles or recognised involvement who identify as women.}
        \label{fig:ratio_scatter}
    \end{minipage}
\end{figure}
Within Belle II, the representation of women as a percentage drops as their career progresses from postgraduate to permanent faculty, shown in Figure \ref{fig:by_pos}. The under-representation of women increases with increasing seniority. 


Figure \ref{fig:ratio_scatter} shows the percentage of women in recognized involvement in the collaboration by year. The trend in this figure shows, again, a small increase over the last 8 years of data collection, but leaves a lot of room for improvement. Internal talks are often given by those in leadership roles such as group chairs. These positions have a low percentage of female representation, therefore the fact that there are even fewer women doing internal talks is not surprising. This figure does show that overall there is an upwards trend, but again this trend is not enough to reach proportional gender representation of our collaboration in the near future.

\subsection{Belle II membership survey 2018}

\begin{figure}[ht]
    \centering
    \begin{minipage}{.45\textwidth}
        \centering
        \includegraphics[width=\linewidth]{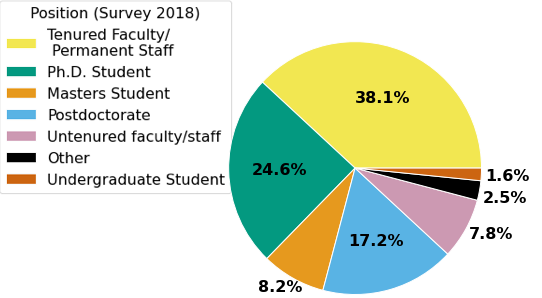}
    \end{minipage}%
    \begin{minipage}{.45\textwidth}
        \includegraphics[width=\linewidth]{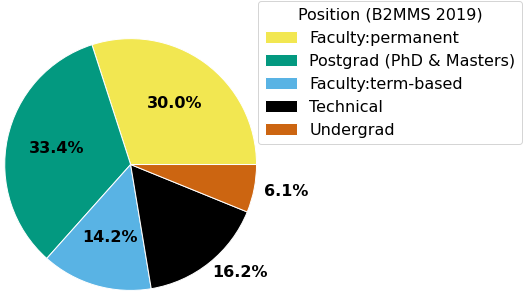}
    \end{minipage}
    \captionsetup{width=.9\linewidth}
    \caption{The collaborators who replied to the Belle II membership survey (left) and the B2MMS 2019 membership (right) by position. The categories recorded are not parallel between the records, therefore there is overlap between some categories.}
    \label{fig:PositionRepresentation}
\end{figure}

A Belle II membership survey for the physicists of the collaboration was conducted in 2018. This survey was inspired by a survey performed by LHCb and is a climate survey asking participants for qualitative and quantitative data. 
It took approximately 6 months to get 244 responses from the collaboration, approximately 30\% of the registered collaboration members at that time. In conducting the poll, the gender representation and career stage was monitored and strong efforts were made to get responses from our junior collaborators. This resulted in an approximately representative sample of responses. Figure \ref{fig:PositionRepresentation} shows the position representations for the 2018 membership survey, compared to the B2MMS FY2019. Due to the nature of surveys - especially optional surveys - sample biasing must be considered whilst analysing the results. Response biasing must also be considered due to the phrasing of the questions and answers.

Belle II learned a lot from the climate survey, including the fact that approximately 25\% of those that took this survey have, at some point in their career, withdrawn from consideration for a leadership role at Belle II (implicitly or explicitly) because of the impact it would have on their family life. This leads to the conclusion that the Belle II collaboration should work on being more inclusive towards family life. 

\section{Belle II diversity and inclusion actions \label{sec:Actions}}

Being such a large, culturally diverse collaboration encourages Belle II to have a code of conduct in which we describe how we expect our members to conduct themselves. Belle II included the first iteration of the code of conduct in the Belle II bylaws in October 2017 and then updated them in 2018. The code of conduct enshrines principles related to research practice and also fostering a diverse and inclusive collaboration and it reads as follows: 
\begin{quote}
    "The Belle II collaboration is committed to fostering an open, diverse, and inclusive working environment that nurtures growth and development of all, and believes that an array of values, interests, experiences, and cultural viewpoints enriches our learning and our workplace. Thus, members shall not engage in violent, harassing, sexist, racist, or discriminatory behaviour." 
\end{quote}
The code of conduct is reiterated at the tri-annual Belle II collaboration meetings during a Diversity plenary and any diversity meetings. 
Alongside the code of conduct, two positions of diversity officer were created in October 2018. Their positions exist:
\begin{itemize}
    \item To promote an inclusive environment within the collaboration;
    \item To provide a safe and confidential point of contact for any collaborator to report any issues, particularly those related to discrimination, bullying, or harassment within the collaboration;
    \item To ensure that persons from marginalised groups are appropriately considered for positions of responsibility in the collaboration and are supported in their careers;
    \item To encourage and publicize the collaboration's events and efforts promoting equity.
\end{itemize}


Belle II uses its active social media platforms to raise awareness of diversity and inclusion events such as International Women’s Day, LGBTSTEM Day, and Colour Blind Awareness Day. The posts are always published in both Japanese and English. 
Belle II aims to always post images that show the diversity that we have within the collaboration and in future plans to publish a series of posts on our collaborators: their position, passions, and experiences. 
In 2019, Belle II became an official supporter of LGBTSTEM Day, through unanimous endorsement of our Institutional Board. Belle II changed its profile picture for the day to a logo with an LBGTQ+ rainbow flag background, and also made a skin available for profile pictures on Facebook. We encourage other research institutions to support each LGBTSTEM day.

In our efforts to improve diversity and inclusion in our collaboration, we must always consider the safety of our members. A collaborator voiced a legitimate concern about having a photo of members alongside the LGBTQ+ rainbow logo in the same post which could be understood as promoting LGBTQ+. This could be detrimental for the safety or situation of participants originating from countries with laws against LGBTQ+ actions. Belle II does not post faces with logos without explicit consent that it is safe for them to do so on social media.

Belle II has been working to improve certain language used in computing and physics. In general, we wish to avoid anything that might cause distress or feelings of exclusion to our collaborators. In particular we are considering words used with severe racial overtones. Our computing and software groups have taken steps to remove and/or phase out the use of the words ‘slave’ and ‘master’ from our code. This has not been straightforward due to external software use, but we have been able to remove `slave' for our build machines for new operating systems and code releases since the beginning of 2020, and we are revisiting ‘master’ now that GitHub is replacing their use of this word. It is now possible for anyone to change personal repositories to use more inclusive language\cite{GitHub}. Belle II is also encouraging alternatives for the terms ‘blacklist/whitelist’.


The planned procedure to record collision data between February and July 2020 at Belle II changed rapidly due to the impact of COVID-19 pandemic. We were able to continue collecting data and still broke the world instantaneous luminosity record thanks to the hard work of the local and remote workers \cite{record}. 
Belle II tried to alleviate some stress caused by the pandemic, the extra work hours and the isolation in the form of snacks and ready meals to show appreciation for shifters and to also increase safety by mitigating the need to go to supermarkets between shifts. We also held some remote social events for example an end of run party on Zoom to celebrate our work and boost moral.


Other important actions worth noting:
\begin{itemize}
 \item Implementing colour blind friendly screens in our control room. Encourage analysts and users to use colour blind friendly colours in plots. All plots in this conference note have used colour blind friendly colour schemes.
 \item The Belle II Secretariat worked to make childcare easier to find as it has been historically difficult to find in Japan.
 \item KEK has improved the bathroom provision for women in the dormitory and is working on improving the situation in experimental areas. There was a request submitted by the International Board of Belle II for a gender neutral barrier free bathroom by our control room; COVID-19 delayed progress but construction started in September 2020.
 \item Belle II members help proofread Japanese to English translations. Often translations or cultural connotations evolve quicker than KEK document revisions and there may be archaic or outdated terms used.
 \item Belle II has asked external food providers for more inclusive food options for dietary restrictions and requirements during collaboration meetings. There is pushback as these meals generally do not sell as well.
\end{itemize}

We will continue to raise awareness within the collaboration through social posts and emails. In particular we wish to normalise making things accessible, for example by encouraging colour blind friendly plots, or recommending to turn cameras on \textit{when speaking} in video conference calls; more of us are experiencing zoom fatigue but video conferencing is particularly trying without facial cues for some with, for example, dyslexia, autism and ADHD.
Finally, we will continue to encourage and normalise putting self care first.

We thank the Belle II Secretariat, Belle II Collaborative Services, and the Belle II Speakers Committee for their invaluable assistance and for maintaining Belle II membership and conference statistics.

\newpage
\bibliographystyle{plain}
\bibliography{references}
\end{document}